\documentclass[useAMS,usenatbib]{mn2e}

\title[Spurs and feathering in spiral galaxies]
{Spurs and feathering in spiral galaxies}
\author[C. L. Dobbs and Ian A. Bonnell]{C. L. Dobbs$^1$\thanks{E-mail:
cld2@st-and.ac.uk} and
I. A. Bonnell$^1$\\
$^1$School of Physics and Astronomy, University of St Andrews, 
North Haugh, St Andrews, Fife, KY16 9SS}
\usepackage{amssymb}
\usepackage{amsmath}
\usepackage{psfig}

\begin{document}

\date{5th February 2006}

\pagerange{\pageref{firstpage}--\pageref{lastpage}} \pubyear{0000}

\maketitle

\label{firstpage}

\begin{abstract}
We present Smoothed Particle Hydrodynamics (SPH) simulations of the response of 
gas discs to a spiral potential. These simulations show that the commonly 
observed spurs and feathering in spiral galaxies can be understood as being 
due to  structures present in the spiral arms that are sheared by the divergent
orbits in a spiral potential. Thus, dense molecular cloud-like structures 
generate the perpendicular spurs as they leave the spiral arms. Subsequent
feathering occurs as spurs are further sheared into weaker parallel
structures as they approach the next spiral passage. 
Self-gravity of the gas is not included in these simulations, stressing that
these features are purely due to the hydrodynamics in spiral shocks.
Instead, a necessary condition for 
this mechanism to work is that the gas need be relatively cold (1000 K or less) 
in order that the shock is sufficient to generate structure in the spiral arms,
and such structure is not subsequently smoothed by the gas pressure. 
\end{abstract}


\section{Introduction}
Understanding the gas dynamics in spiral galaxies is an important precondition 
if we are to comprehend the physics of star formation
(\citet{Elmegreen2002, Elmegreen1999, Bonnell2006}) and galaxy evolution. 
The origin of spurs and feathering is one phenomenon that has proved difficult
to explain. These
features are commonly observed in many spiral galaxies, including grand design
galaxies such as M51 (\citet{Aalto1999}). Spurs (particularly in M51)
often show a correlation with HII regions, suggesting a link between these
structures and star forming giant molecular clouds. In fact the local
star-forming region of Orion has been referred to as the Orion spur
(Bok1959). Therefore, understanding the
origin of spurs may offer insights into the star formation process.
 
Spurs are seen as narrow dark lanes that extend perpendicular to the spiral arms
\citep{Lynds1970, Weaver1970}. They are particularly striking in the HST images of M51
\citep{Scoville2001}, especially towards the
centre of the galaxy. The spurs extend over half the distance between the arms, 
and are often dotted with HII regions.
Surveys indicate that spurs
have pitch angles of $30-50^o$ \citep{Russell1992} and widths comparable to 
those of spiral arms \citep{Elmegreen1980}.
Further examples of spurs are also shown in \citet{Byrd1983} (M31) and 
\citet{Kaufman1989} (M81).

We denote feathering \citep{Balbus1988, Kim2002} by parallel filamentary features 
shifted from the spiral arms. These are often observed on the outskirts of 
spiral galaxies (e.g. M74).
Branches are more significant longer spurs often associated with resonances. 
Several simulations already have shown the bifurcation of spiral arms primarily 
at the 4:1 
resonance of their potential \citep{Patsis1994,Patsis1997,Chak2003}.     
 
Spurs have been suggested to form mainly through gravitational or magneto-Jeans instabilities.  
\citet{Balbus1988} demonstrated that transient gravitational stabilities develop
preferentially nearly parallel and nearly perpendicular to the spiral arm. This introduces the 
appealing scenario where spurs and density perturbations along the arms are generated by the 
same mechanism.  
Generally, dense regions along the course of the spiral arm will 
produce projections roughly perpendicular to the arm as a result of shear
expansion. Simulations by \citet{Kim2002} show the formation and
fragmentation of spurs in 2D magneto-hydrodynamical calculations of a shearing
box. Dense clumps are produced in the arms principally through the magneto-Jeans 
instability, leading to the formation of spurs.   

Previous numerical analysis by \citet{Dwark1996} 
indicated that galactic flows through spiral arms are stable against purely 
hydrodynamic instabilities.
However \citet{Wada2004} have looked again at the  
non-self-gravitating, non-magnetic case, performing global simulations of a
spiral galaxy. They observe a rippling distortion of the shock front and the
formation of spurs at larger pitch angles. These spurs are interpreted as a
consequence of Kelvin-Helmholtz instabilities. 

We present hydrodynamic simulations which show the development of feathering
and spurs in a spiral galaxy. These features
are much more evident in our results than previous purely hydrodynamical
simulations.
We note that most previous studies, including \citet{Wada2004} and 
\citet{Dwark1996}, assume a high sound speed.
Our analysis reveals that the temperature of the disk has a 
crucial effect on the disk structure and the formation of spurs. We show that
these features are due to the shearing of structure in the spiral arms.
Furthermore, inhomogeneities transmitted between the arms lead to the 
formation of new structure in the next arm.
       
\section{Calculations}
We use the 3D smoothed particle hydrodynamics (SPH) code based on the version by
Benz \citep{Benz1990}. 
The smoothing length is allowed to vary with space and
time, with the constraint that the typical number of neighbours for each 
particle is kept near $N_{neigh} \thicksim50$.
Artificial viscosity is included with the standard parameters $\alpha=1$
and $\beta=2$ \citep{Monaghan1985,Monaghan1992}.

\subsection{Flow through galactic potential}
The galactic potential includes a 4 armed spiral component, from \citet{Cox2002}.
The symmetric components consists of a logarithmic
potential (e.g. \citet{Binney}) that provides a flat rotation curve of
$v_0=220$~km~s$^{-1}$, and a potential for the outer halo \citep{Cald1981}.  
Parameters for the spiral part include the amplitude, 1~atom~cm$^{-3}$,
pattern speed, $2 \times 10^{-8}$~rad~yr$^{-1}$ and pitch angle $\alpha =
15^o$ which are comparable with 
the Milky Way. The pattern speed leads to a co-rotation radius of 11kpc. 
 
Overall the disk is in equilibrium, as the rotational velocities of disk particles balance the centrifugal force from
the potential. This paper only considers how hydrodynamic forces and galactic
potential influence the flow. In particular, self-gravity magnetic fields,
heating, cooling or feedback from star formation are not included.
 
\subsection{Initial conditions}
Gas particles are initially chosen to occupy a region of radius
5 kpc$<r<$10 kpc. The disk also has a scale height $z\leq 100$ pc.  
Particles are allocated positions and velocities determined from a 2D test
particle run. The test particle run consists of 1 million particles 
initially distributed uniformly with
circular velocities. They evolve for a couple of orbits subject to the
galactic potential, to give a spiral density pattern with particles settled into
their perturbed orbits. For the highest resolution run (4 million particles), 
$4.5 \times 10^5$ test particles were split into 9 particles each.

In the SPH initial conditions, the particles are given velocities in the z
direction from a random Gaussian distribution of 2.5\% of the orbital speed.
The same magnitude velocity dispersion is also added to the x and y velocities.
The gas is distributed uniformly on large scales with an 
average surface density of $\Sigma
\approx 2$ M$_{\odot}$ pc$^{-2}$, although it
is somewhat clumpy on smaller scales. The densities in the interarm 
regions range from $10^{-25}$ g cm$^{-3}$ to $10^{-24}$ g cm$^{-3}$.
The total mass of the disk is $5\times10^8$ M$_{\odot}$.

The number of particles is either $10^6$ or $4\times10^6$. 
The higher resolution work was
carried out on UKAFF (UK Astrophysical Fluids Facility). Simulations were run for
between 200 and 300 Myr, where the orbital period at 5 kpc is 150 Myr. 
No boundary conditions are applied to our calculations. 

All calculations are isothermal with temperatures of 50, $10^2$, $10^3$, or 
$10^4$ K. Most previous simulations and analysis of spiral galaxies 
(e.g. \citet{Wada2004,Gittins2004,Kim2002,Dwark1996}) assume a sound speed of 
approximately 
10 km s$^{-1}$ corresponding to the hotter (10$^4$ K) component of the ISM.
However, for GMCs to form, gas entering the spiral arms must be cold atomic
clouds \citep{Elmegreen2002} or pre-existing molecular gas \citep{Pringle2001}.
Since our initial aim of this work was to investigate molecular cloud
formation (described in Dobbs et. al. 2006), lower ISM temperatures were adopted for these
calculations. 
 
\section{Results}

The highest resolution run used $4 \times 10^6$ particles, giving a resolution of
125 M$_{\odot}$ per particle. The temperature in this simulation was 50 K. 
Density plots at 4 different times are shown in Fig.~1. These and subsequent
figures are shown in the rotating frame of the spiral potential.

Initially there is a smooth spiral perturbation to an approximately uniform
disk. 
For the 50 K gas, the spiral arms are initially very thin. The density of the
spiral arms increases with time and non-uniform structure develops along each 
arm. 
As this arm substructure grows, dense clumps shearing away from the spiral arms
lead to interarm features. These features occur on the leading side of the spiral
arm, leaving the trailing side smooth.
Structure on the outside of spiral arms is noticeable first in the inner 
parts of the disk (Fig. 1b), and then spreads to larger radii. This is not surprising since the
orbital period is less for smaller radii, so the comparative evolution of that
part of the disk is faster.  
Both the arm and interarm regions become more
disordered with time, and consequently the spiral arms become wider.

Further runs have been performed at 100, $10^3$ and $10^4$ K 
with 1 million particles and are shown in Fig.~2. 
As the temperature increases, the Mach number of the shock decreases and much
weaker spiral arms are produced. The gas is largely smooth at $10^4$ K
due to the higher pressure, and the spiral arms exhibit no obvious substructure.

\begin{figure*}
\centering
\begin{tabular}{c c}
\psfig{file=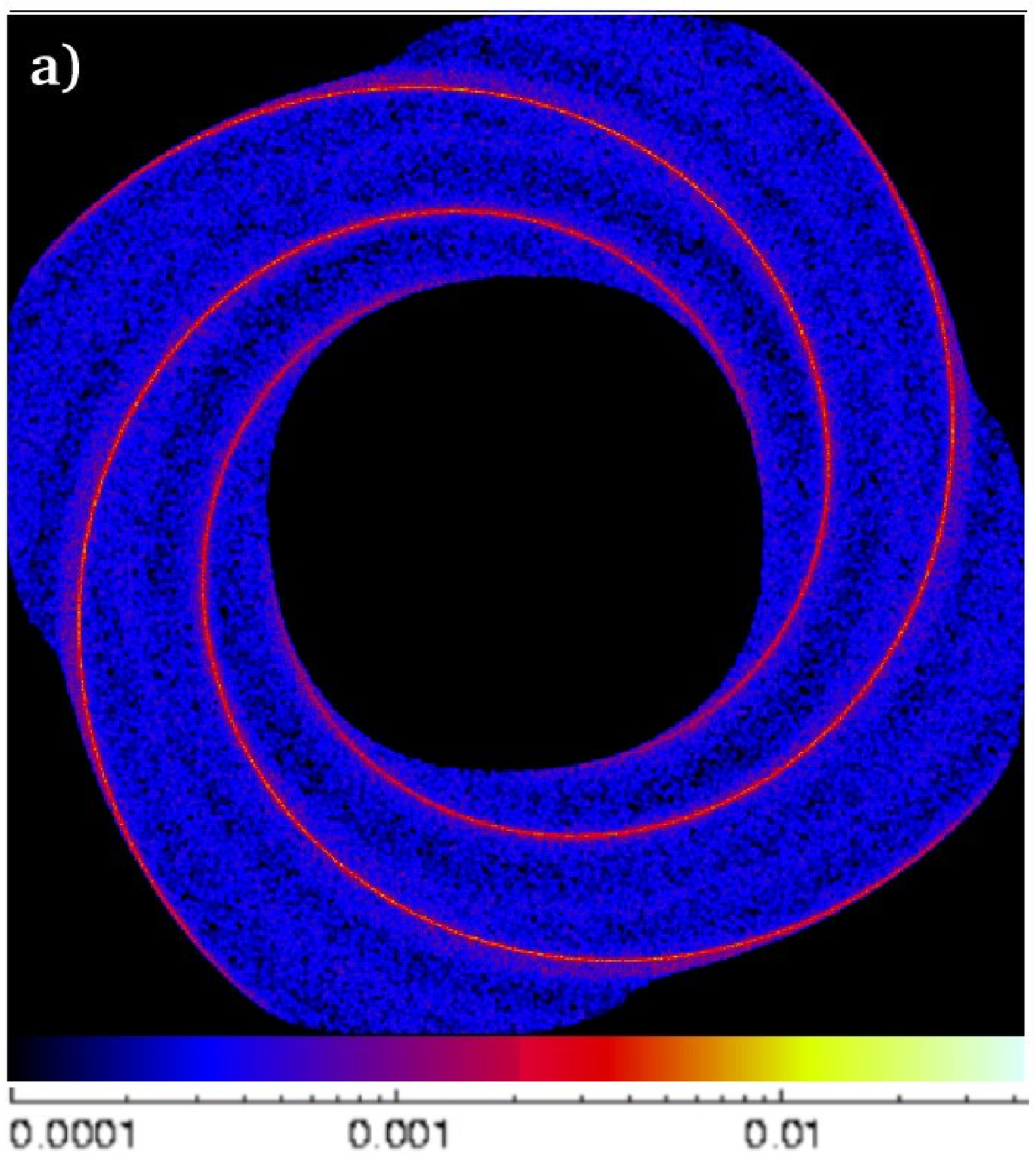,height=2.8in} & 
\psfig{file=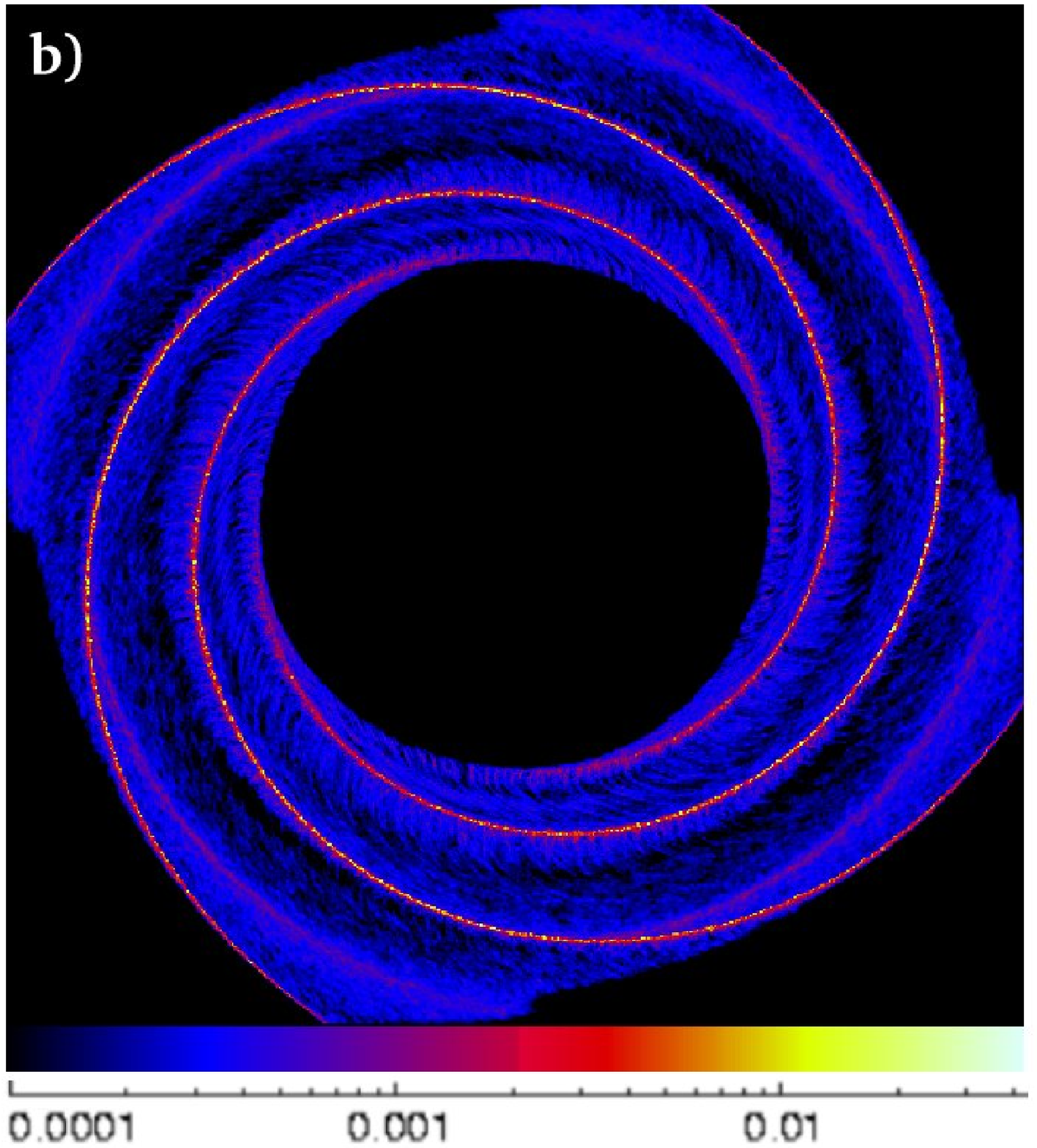,height=2.8in} \\ 
\psfig{file=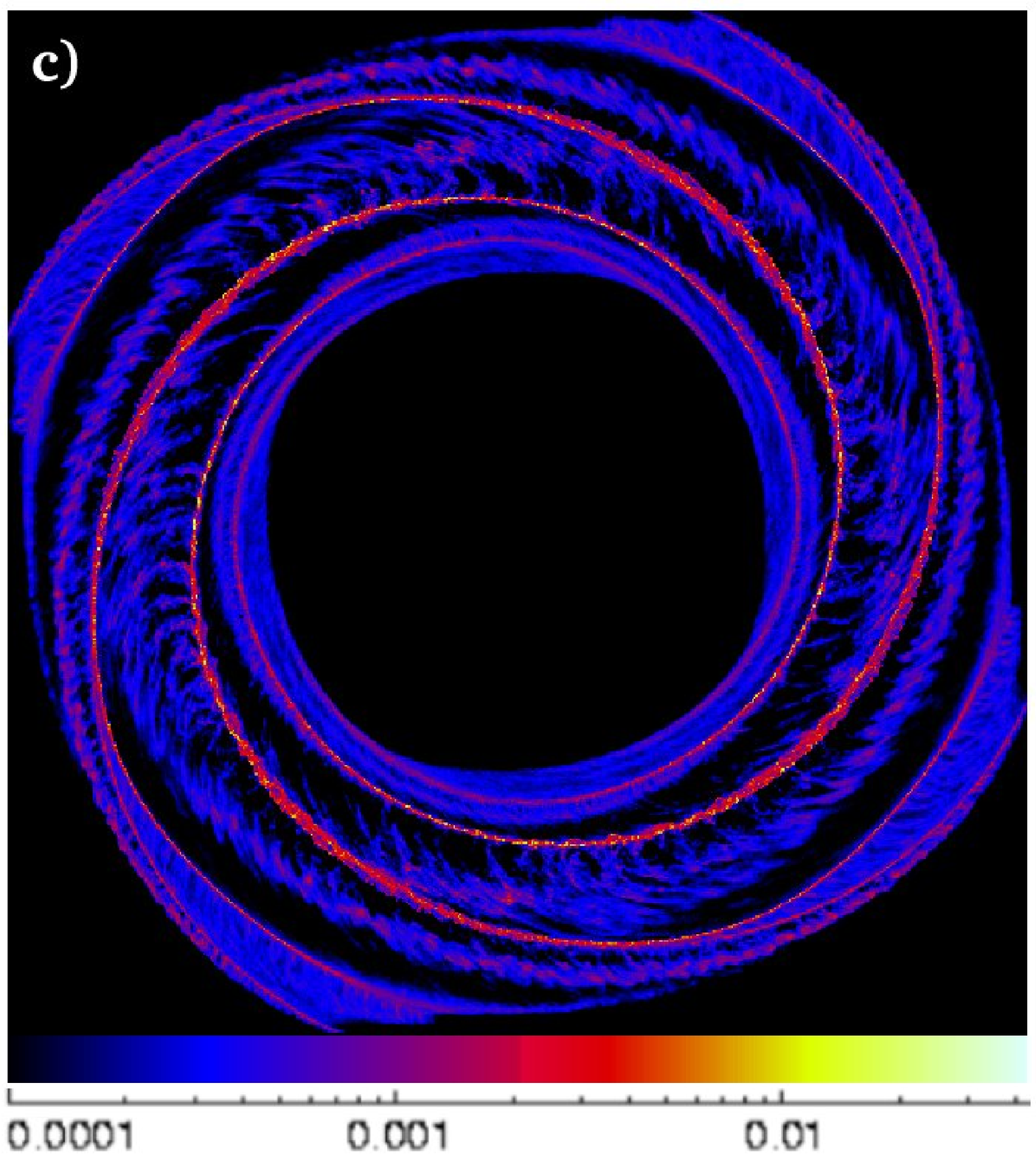,height=2.8in} &
\psfig{file=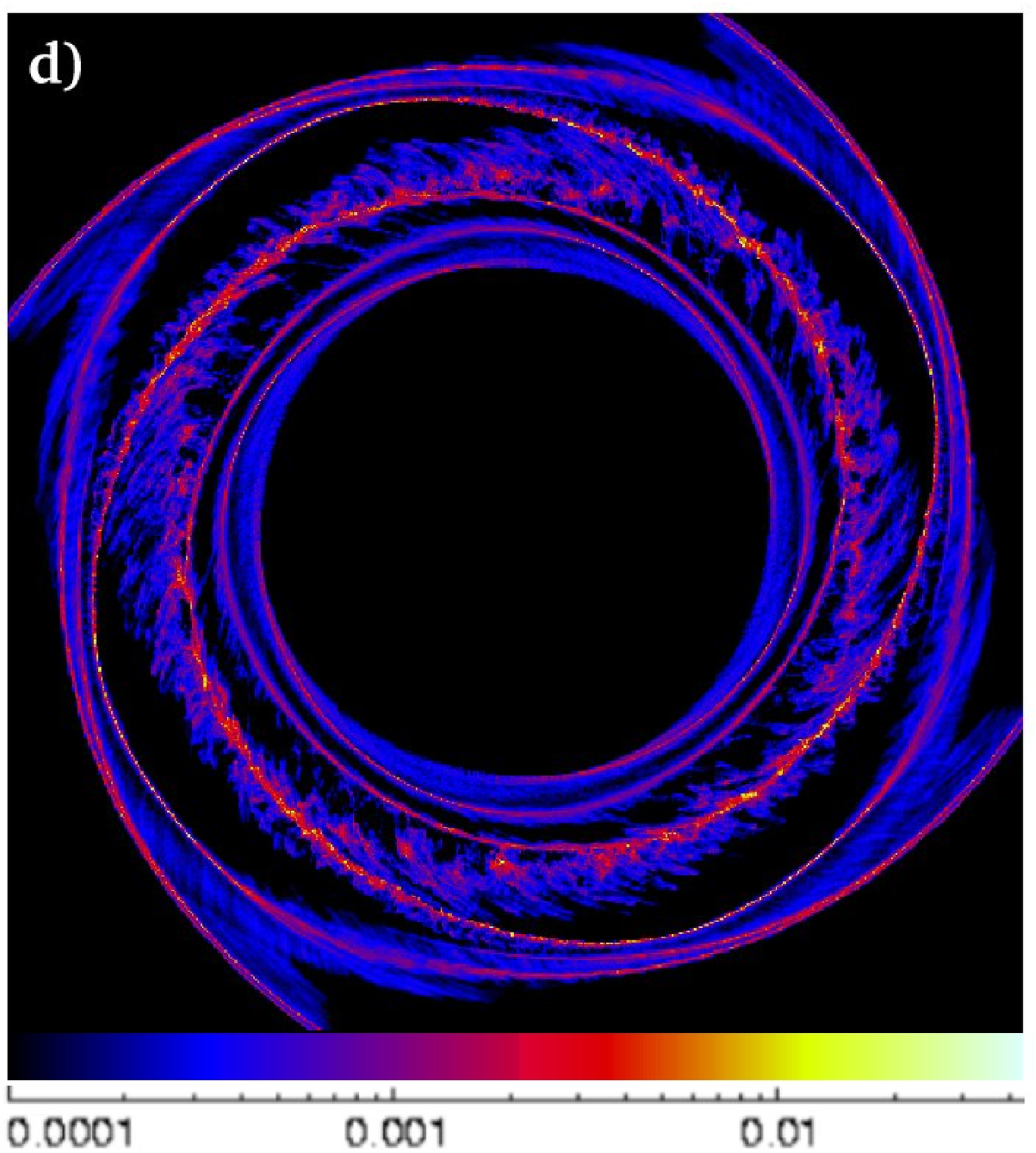,height=2.8in} \\
\end{tabular}
\caption{Column density plots (g cm$^{-2}$) 
when T=50 K after a) 0, b) 60, c) 160 and d) 260 Myr.
The number of particles is 4 million and each plot is 20 kpc by 20 kpc. 
The same scaling is used on all subsequent
column density plots.}
\end{figure*}

\begin{figure*}
\centering
\begin{tabular}{c c c}
\psfig{file=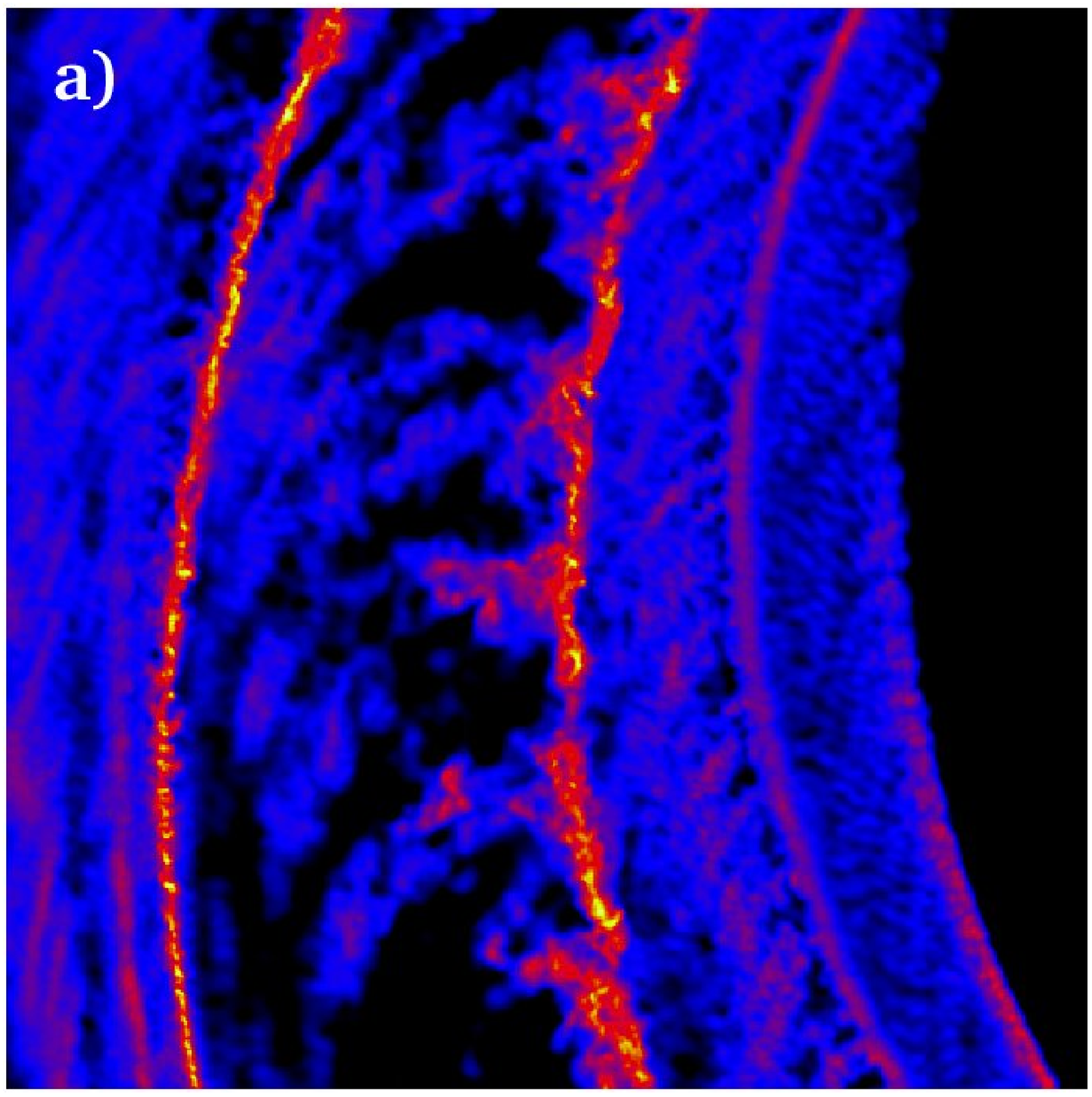,height=2.2in} & 
\psfig{file=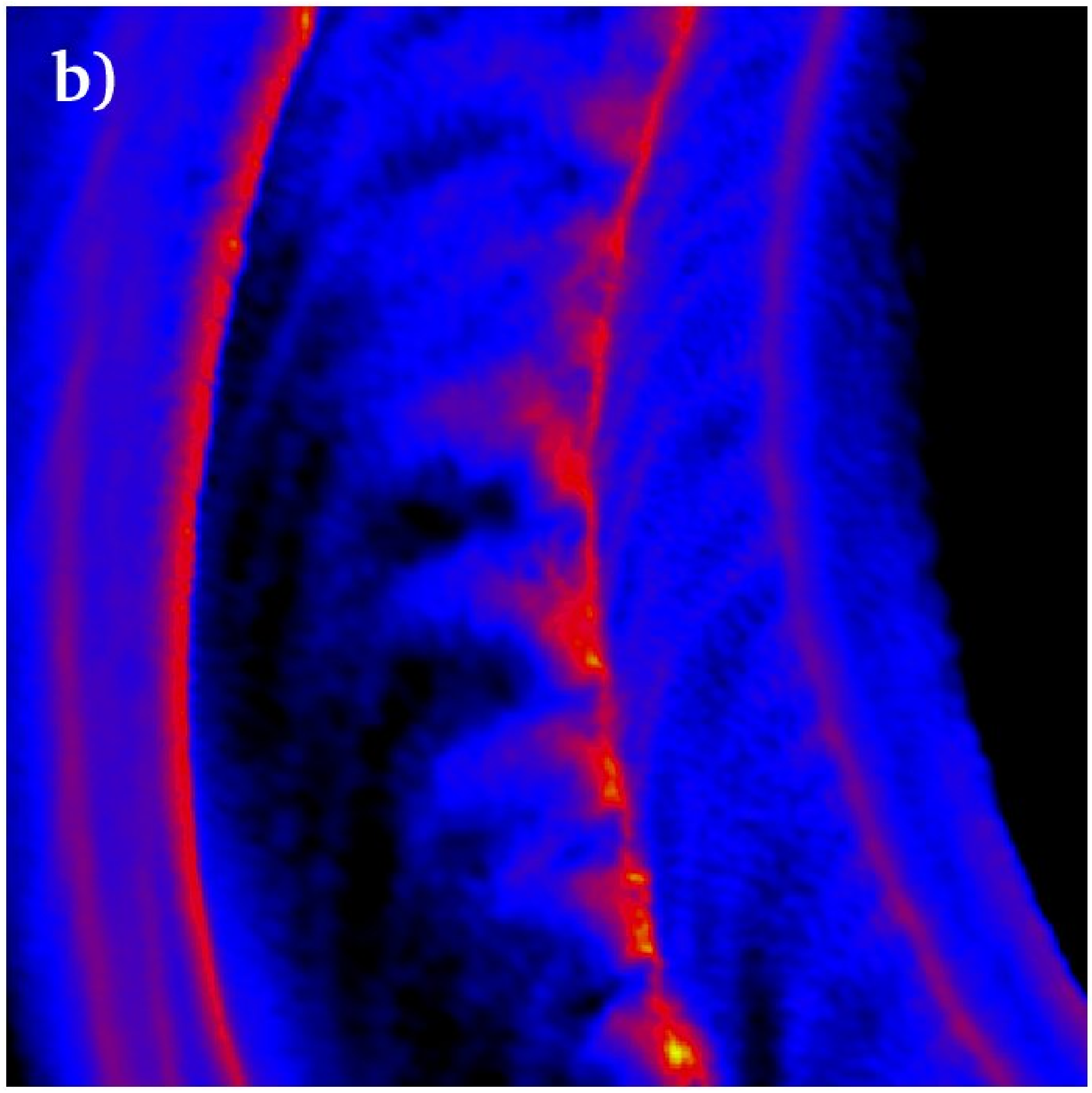,height=2.2in} &
\psfig{file=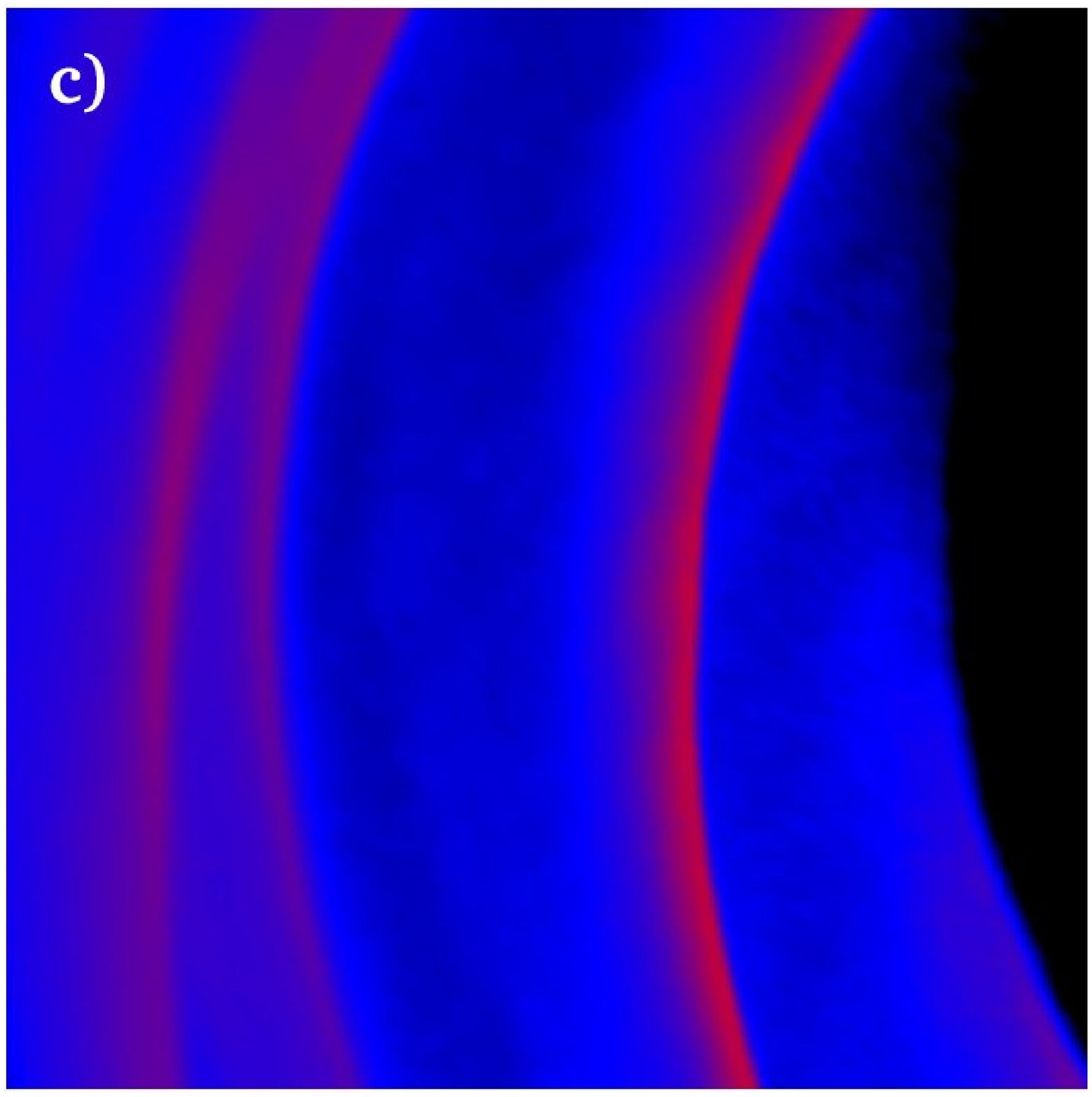,height=2.2in} \\
\end{tabular}
\caption{Column density plots (g cm$^{-2}$) when
the temperature is a) 100, b) $10^3$ and c) $10^4$ K. The time corresponding for
each figure is 220 Myr. Each plot shows a 5 kpc by 5 kpc
section of the global simulation. The overall number of particles in these
simulations is 1 million.}
\end{figure*}
The detailed structure of the spurs and feathering in the 50 K simulation
is shown in Fig.~3.  We see that the interarm structure increases with time
from smaller scale
structures to more pronounced larger scale spurs and feathering. 
Over time, spurs become more distinct with a regular spacing 
($\approx 0.7$ kpc after 160 Myr in Fig.~3c). 
Furthermore, we see that
there is a direct correlation between the spurs that leave perpendicular to the spiral arm and
the feathering that approaches the next spiral arm. 
The feathering
here has the appearance of closely spaced parallel lanes of gas with a much 
smaller pitch angle than apparent for spurs.
The interarm structure in 
the 1000 K run is also shown in Fig.~2b. The spurs, and to a lesser extent the feathering, are still
evident but are less pronounced. As noted above, the $10^4$~K simulation does not display such interarm structure,
but instead retains a smooth gas distribution (Fig.~2c).


\subsection{Formation of spurs}
In order to ascertain the origin of the interarm structure, we have traced the gas backwards in time from the spur
that leaves the spiral arm to when the gas was located in the spiral arm itself.
Figs.~4 and 5 show that there is
a direct correlation between clumps in the spiral arms and the spurs that leave perpendicular to the arms.
Thus we can conclude that the spurs are  simply  the shearing of clumps in the 
spiral arms, due to the divergent orbits of material that leave the arm.
As structure (GMCs) is observed to exist in spiral arms,
this provides a simple explanation for the presence of spurs.
Further shearing then produces the feathering from the spurs.
A similar scenario is discussed in \citet{Gittins2004}, who also 
describes the formation of spurs in SPH calculations.  

The clumpy structure of the spiral arms is due to the dynamics as the ISM passes
through a spiral shock \citep{Dobbs2006}. We interpret the growth of
clumpy structure in the spiral arms in terms of the particles' change in
angular momenta in the shock, which modifies the velocity phase space 
distribution of particles in the disk.
The particles enter the shock at non-uniform intervals of space or time   
(since the interarm regions are not homogeneous). As they
subsequently gain and/or lose angular momentum in the spiral arms, the particles 
tend to group together in velocity space \citep{Dobbs2006}.
Hence the inhomogeneities of the initial gas distribution are
amplified in the spiral arms and clumps form. 
We estimate the spacing of these clumps from the time particles spend in the 
spiral arm ($t_{arm} \sim 6 \times 10^7$ 
years) and the typical velocity spread ($v_{\perp} \sim 10$~km s$^{-1}$) 
\citep{Dobbs2006} to be $L \sim t_{arm} v_{\perp} \sim 600$~pc.

We plot the paths of five particles and their corresponding angular momenta in
Fig.~6. Since the spiral arms are regions of significantly enhanced density, gas
particles from a range of radii must interact in the shock. Particles entering
the shock on epicyclic paths are typically at the furthest extent of their
orbit. They therefore encounter higher angular momentum material already in the
shock (e.g. red line, Fig.~6). This generally leads to a jump in the angular
momentum of the particles as they enter the shock (e.g. yellow line, Fig.~6).  
As particles travel along the spiral arm, they tend to lose angular momentum, 
since they are constrained by the spiral potential. 
It appears in Fig.~6 that particles entering the shock later (e.g. green line) 
show less change in angular momentum, possibly because they interact with gas 
in the shock which has similar angular momentum. Hence the path represented by 
the green line is less affected by the shock, and this particle spends less 
time in phase with the potential compared to the others. 
Eventually the gas leaves the spiral arm when it's angular momentum is too 
high to travel further radially inwards, and forms the spur shown in Figure~4..

This model also explains why less structure is apparent in the higher
temperature simulations. At higher ISM temperatures, the shocks are weaker and 
they have less effect on the particles' angular momentum. The particles spend a
shorter time in phase with the potential and their orbits are less perturbed.
Overall the distribution of the angular momenta of particles remains much more 
uniform \citep{Dobbs2006}.   
Furthermore, the gas pressure can smooth out much of the structure 
during the crossing of the spiral arm. If the structure in the spiral
arms is removed, then there is nothing  to shear into the subsequent spurs and 
feathering in the interarm region.
For example, the $10^4$ K run has a sound speed of $\approx$~10~km~s$^{-1}$ which
is comparable 
to the velocity of the gas perpendicular to the
spiral arm (in the rotating spiral potential). Thus, structures of size 
scales $\le$ the width of the arm can be smoothed by the
internal pressure before leaving the spiral arm.

An alternative explanation is that the spiral arm structure arises
from Kelvin Helmholtz instabilities \citep{Wada2004}. We find that the Reynolds
numbers for our simulations are relatively low ($Re\approx 500$ for 100 K gas
due to the numerical resolution)
implying that the resolution is insufficient for Kelvin Helmholtz instabilities
to occur.
Interestingly, \citet{Wada2004} find spurs forming from hot ($10^4$)
gas, although their analysis likewise shows that increasing the Mach number of 
the shock (corresponding to a lower temperature) increases the susceptibility of
gas disks to hydrodynamic instabilities.
\begin{figure*}
\centering
\begin{tabular}{c c c}
\psfig{file=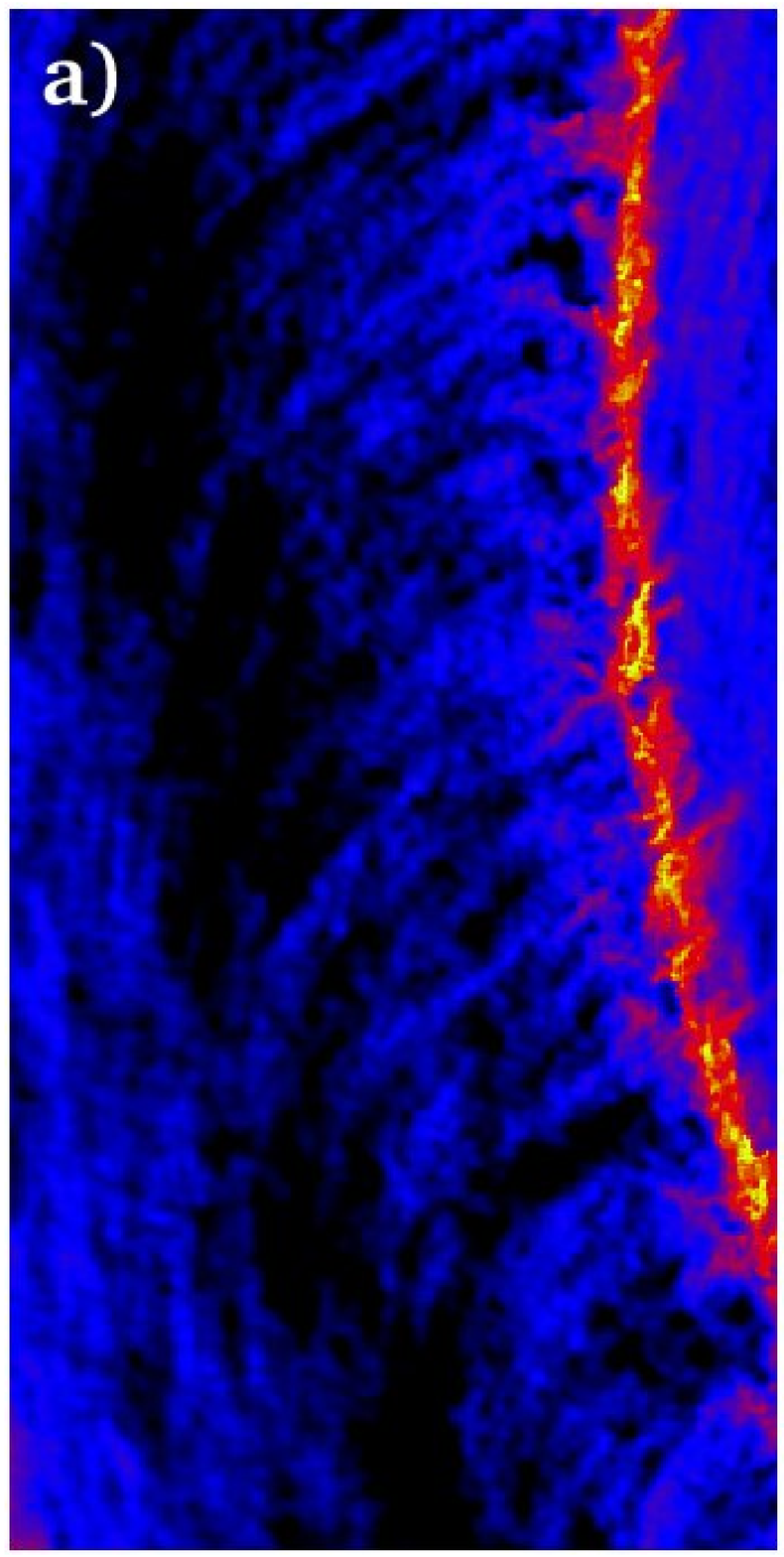,height=2.8in} & 
\psfig{file=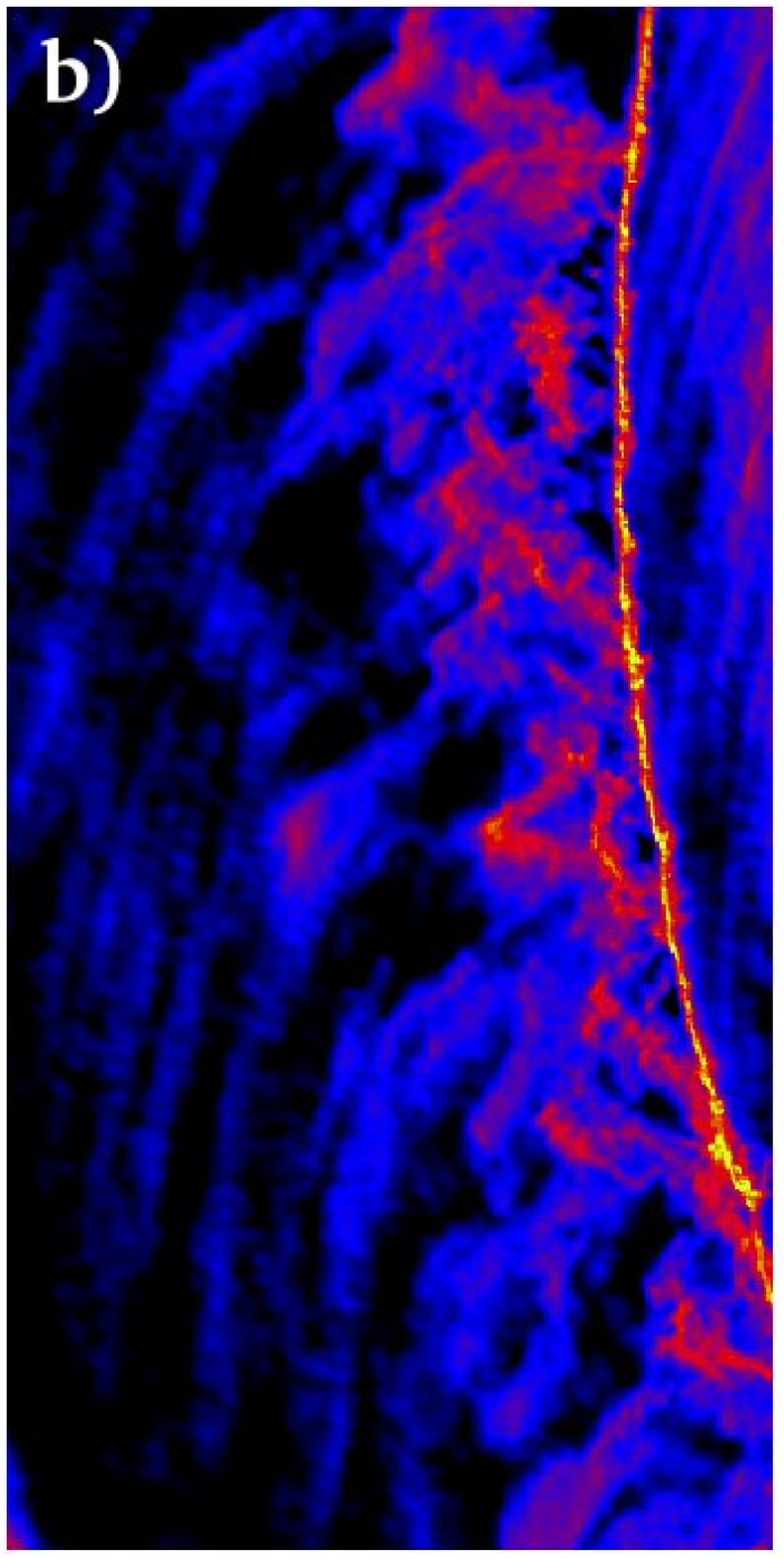,height=2.8in} &
\psfig{file=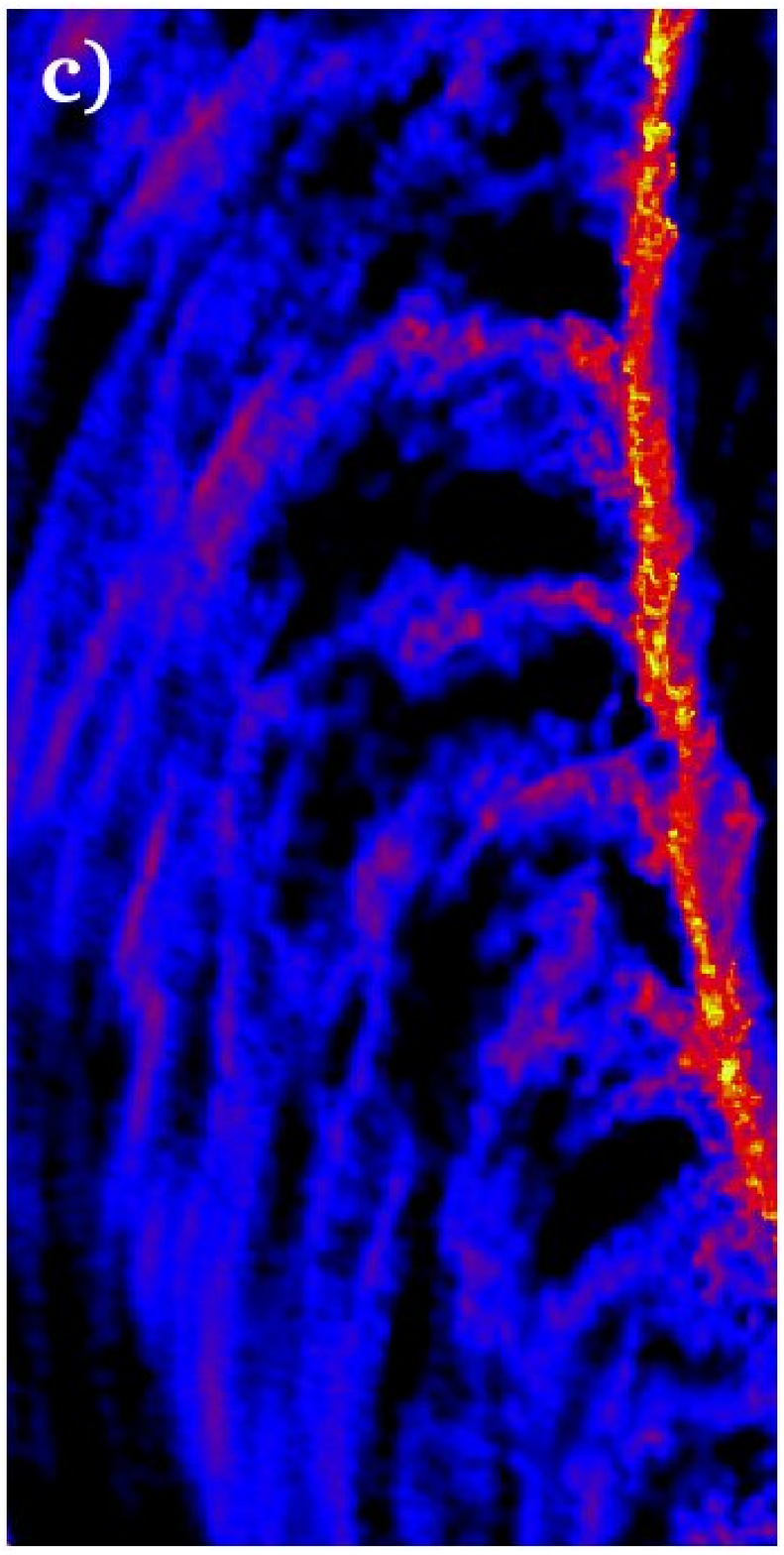,height=2.8in} \\
\end{tabular}
\caption{Time evolution of feathering and spurs. 
Column density plots for the 50 K run after a) 80 Myr, b) 120 Myr and c) 160 Myr.
Each plot is 2 kpc by 4 kpc.}
\end{figure*}

\begin{figure*}
\centering
\begin{tabular}{c c}
\psfig{file=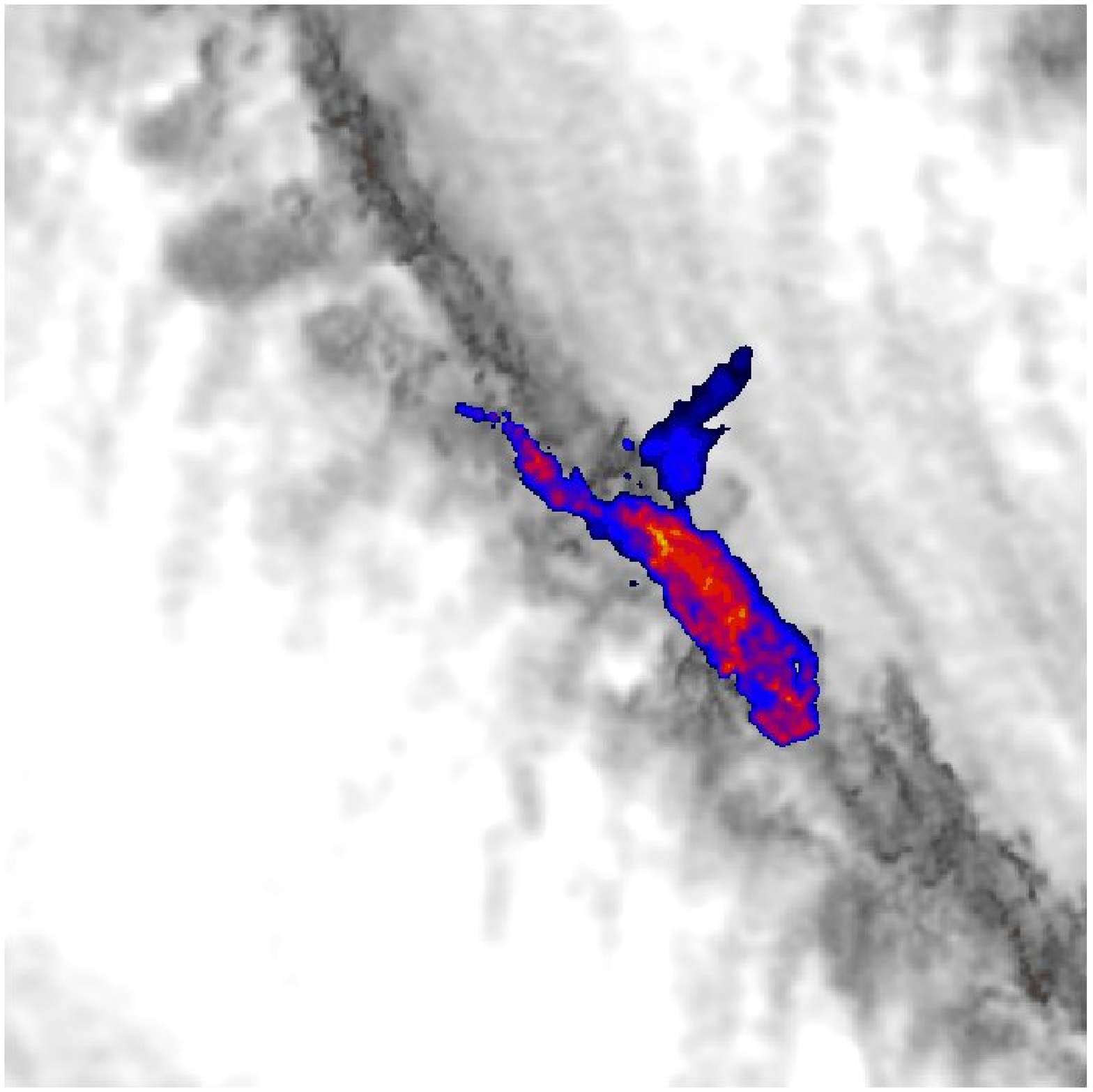,height=2.3in} & 
\psfig{file=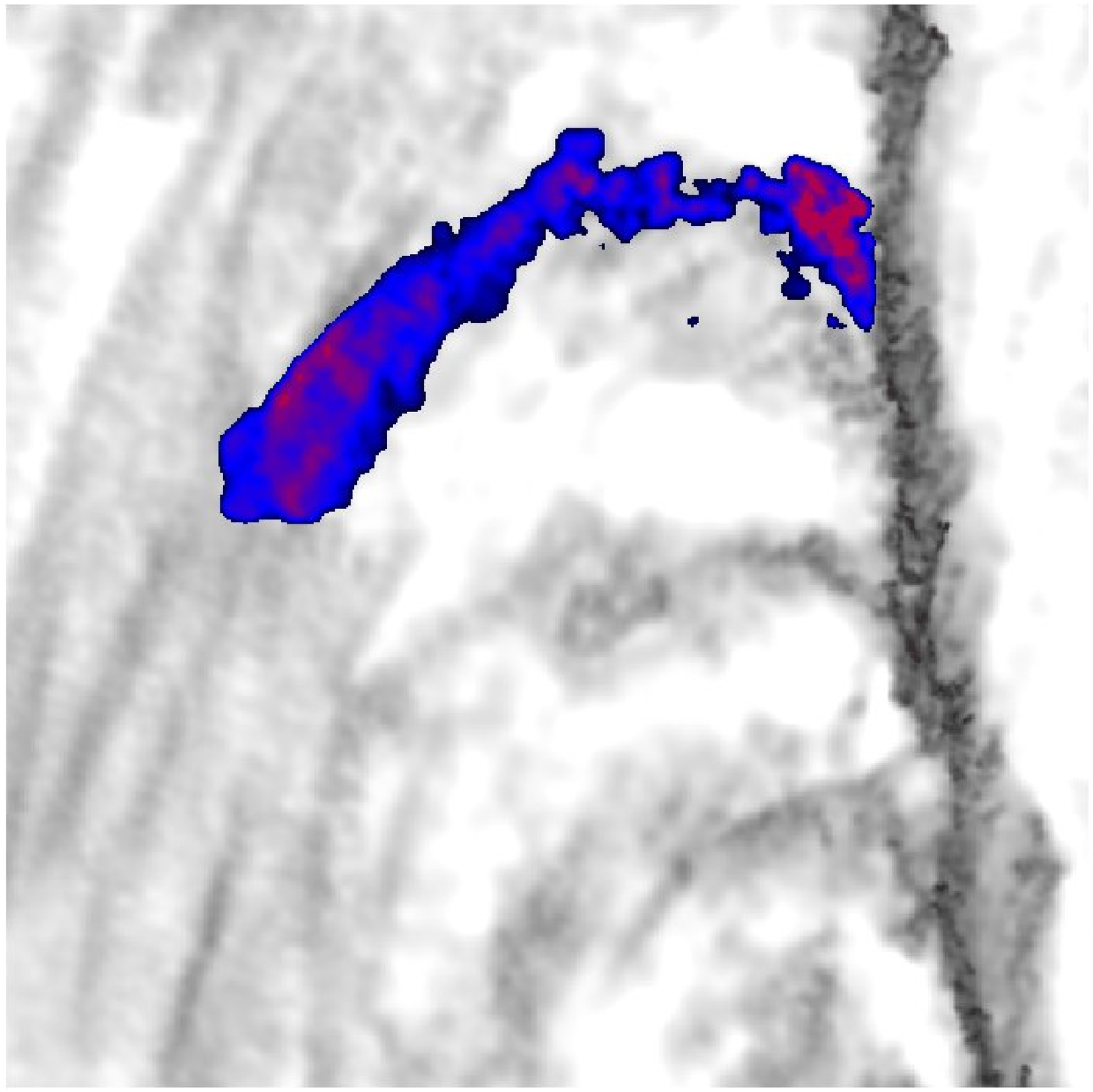,height=2.3in} \\
\end{tabular}
\caption{Formation of a spur (right hand plot, after 160 Myr) from a dense clump of gas in
the spiral arm (left hand plot, after 100 Myr). The temperature is 50 K and both
plots are 2 kpc by 2 kpc.}
\end{figure*}

\begin{figure*}
\centering
\begin{tabular}{c c}
\psfig{file=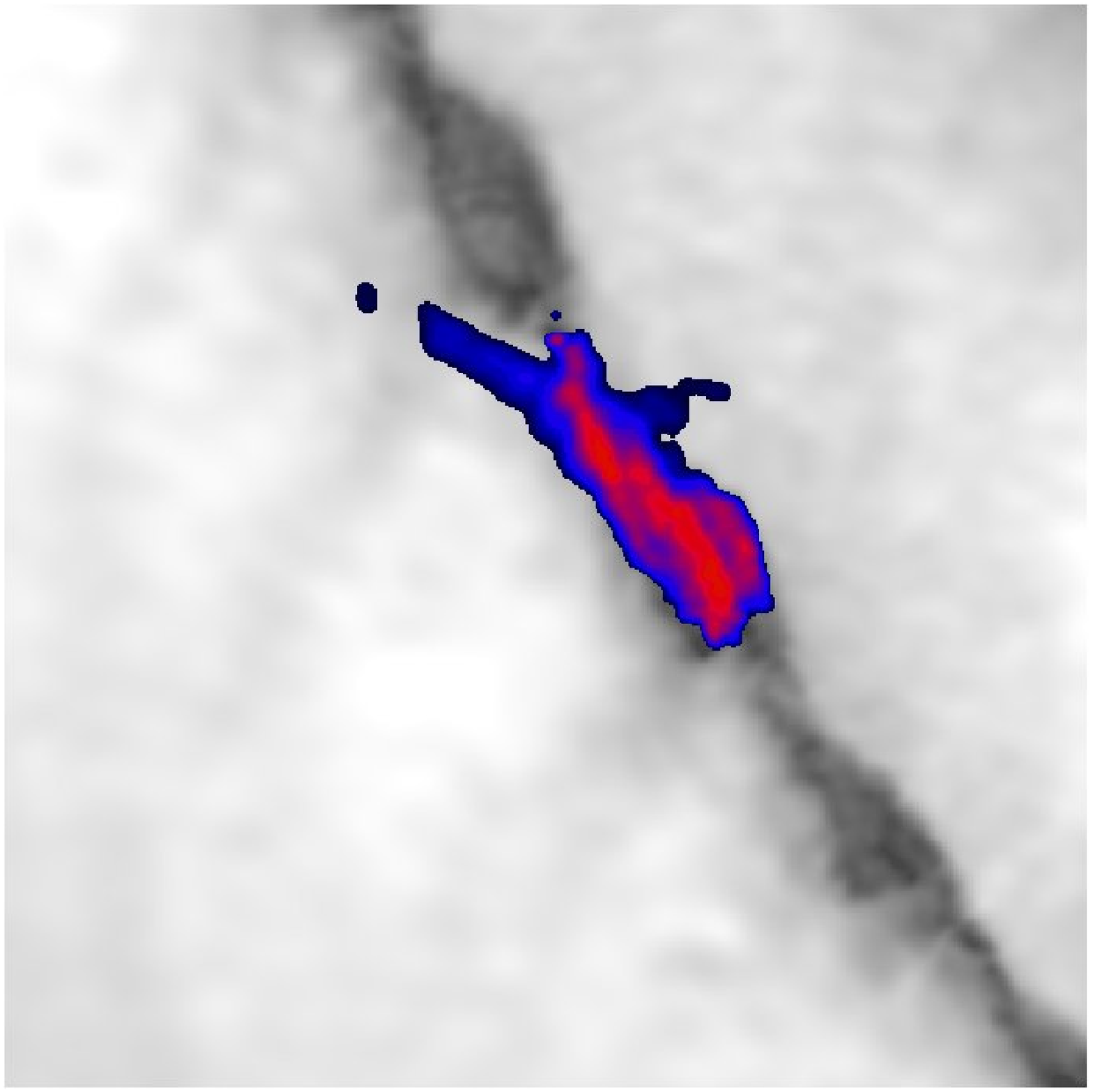,height=2.3in} & 
\psfig{file=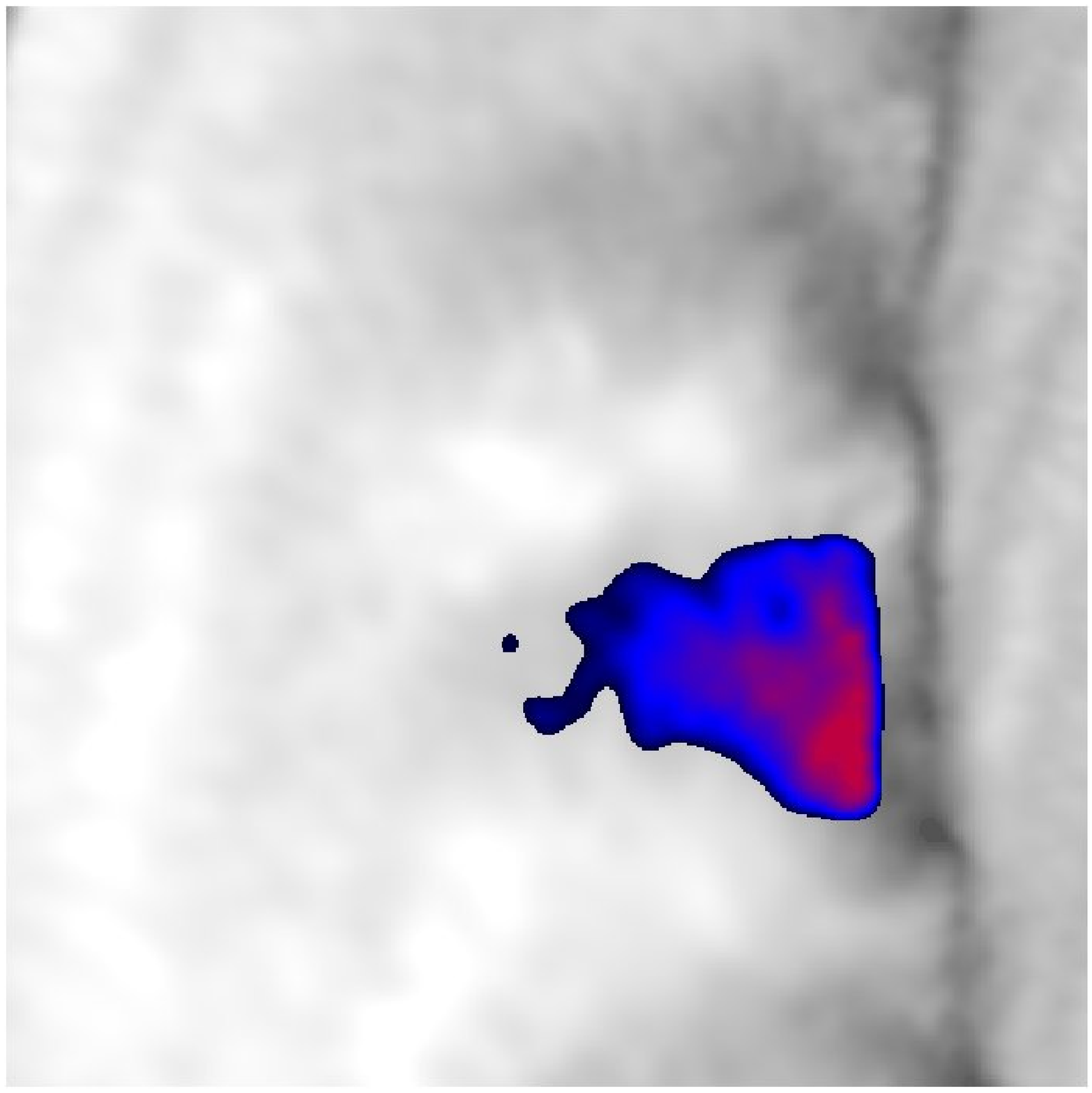,height=2.3in} \\
\end{tabular}
\caption{Formation of a spur (right hand plot, after 220 Myr) from a dense clump of gas in
the spiral arm (left hand plot, after 180 Myr). The temperature is 1000 K
and both plots are 2 kpc by 2 kpc.}
\end{figure*}

\begin{figure}
\centering
\begin{tabular}{c}
\psfig{file=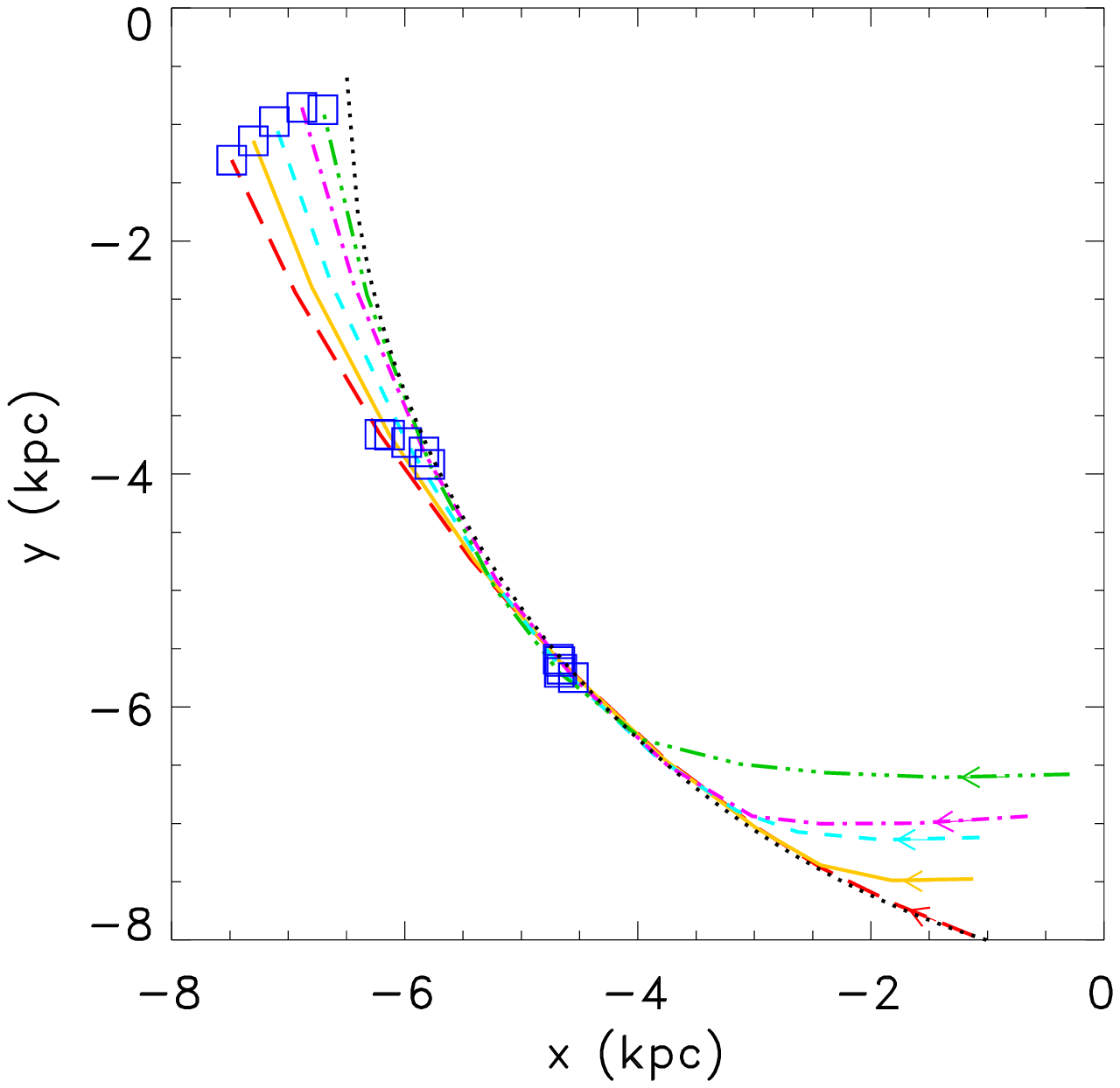,height=3in} \\
\psfig{file=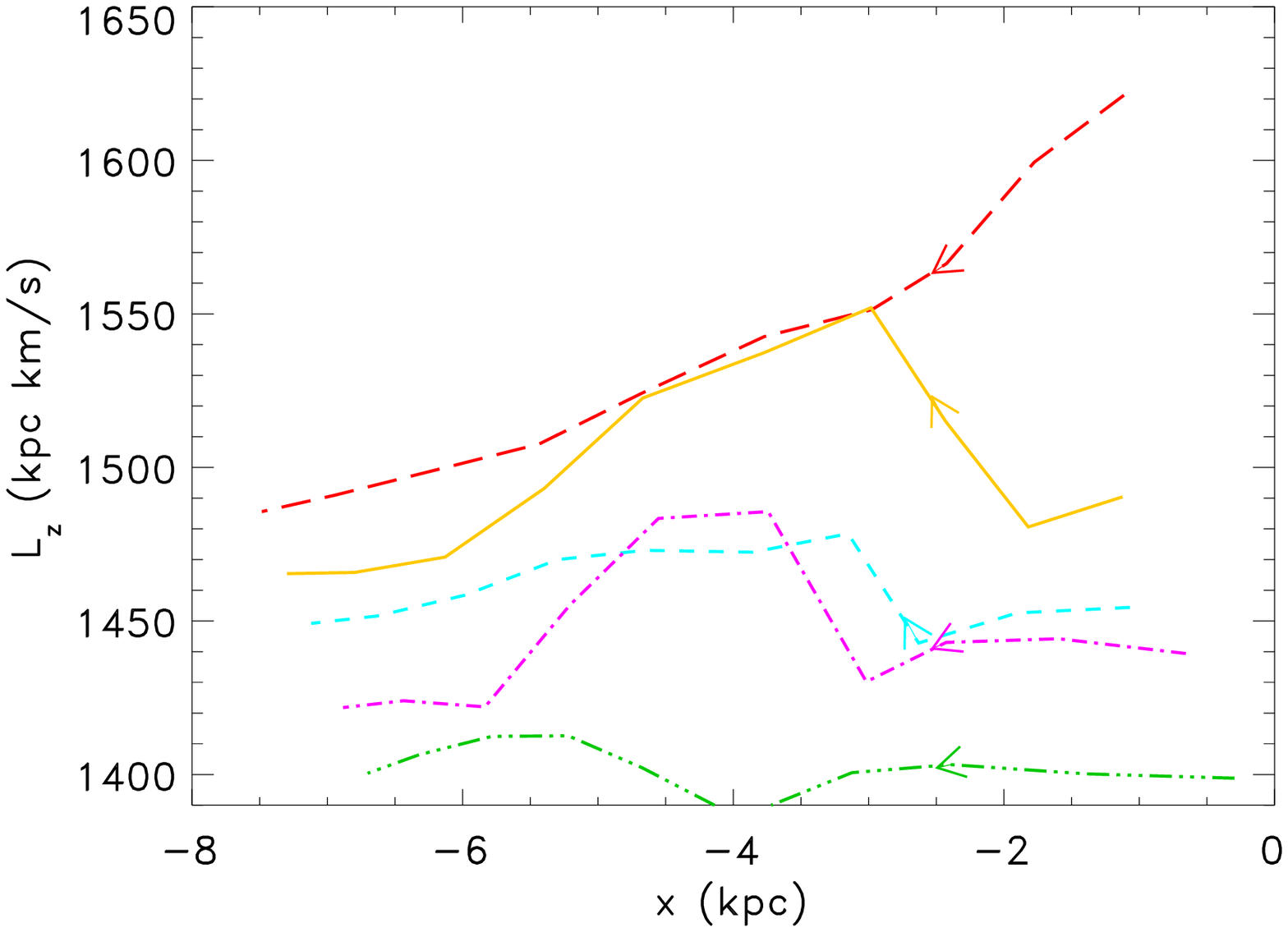,height=2.4in}
\end{tabular}
\caption{The orbits of 5 selected particles are displayed pre- and post-shock
(top) where the xy-coordinates correspond to a fixed Cartesian grid centred at the
midpoint of the disk. The dotted line
indicates the position of the shock front, and the orbits are taken from the
rest frame of the potential. Boxes indicate the location of the five particles
at three corresponding times. The final position of the particles coincide with
the spur shown in Figure~4 (right). The corresponding angular momenta of the
particles are also plotted (bottom). Particles are travelling from right to left
in both plots.}
\end{figure}
 
\section{Conclusion}
We have performed numerical simulations of disks in spiral galaxies which
demonstrate 
the widespread formation of spurs and feathering. 
The overall appearance of the galactic disk, enhanced by these features, is 
remarkably similar to observations of galaxies such as M51 and M81.
In addition, the properties of individual spurs (pitch angle, dimensions) 
are comparable with observations. 

The formation of spurs and feathering in the interarm regions of spiral galaxies
is shown to be linked to the presence of structure in the spiral arms. These
findings concur with similar work by \citet{Gittins2004}.
Upon leaving the spiral arms, dense concentrations of gas are sheared by
diverging orbits. Spurs form perpendicular to the spiral arms, which later
become feathering that impacts into the next spiral arm. 
The formation of spurs and GMCs are thus directly
related, since these denser structures in the spiral arms are the sites of 
molecular clouds.

The necessary 
conditions for interarm features to form are
density inhomogeneities and relatively cold gas (T $\lesssim$ 1000 K). 
A clumpy ISM allows structure growth in cold
spiral shocks. Such structure formation in the spiral arms, and therefore spur
formation, does not occur in our simulations when the gas is hot.     

\section*{Acknowledgements}
Computations included in this paper were performed using the UK Astrophysical
Fluids Facility (UKAFF).

\bibliographystyle{mn2e}
\bibliography{Dobbs}

\bsp

\label{lastpage}

\end{document}